\documentclass{article} 

\usepackage{arxiv} 

\usepackage[utf8]{inputenc} 
\usepackage[T1]{fontenc}    
\usepackage{url}            
\usepackage{booktabs}       
\usepackage{amsfonts}       
\usepackage{nicefrac}       
\usepackage{microtype}      
\usepackage{lipsum}         
\usepackage{graphicx}
\usepackage{doi}
\usepackage{authblk}
\usepackage{amsmath,amssymb,amsfonts}%
\usepackage{multirow}%
\usepackage{amsthm}%
\usepackage{mathrsfs}%
\usepackage[title]{appendix}%
\usepackage{xcolor}%
\usepackage{textcomp}%
\usepackage{manyfoot}%
\usepackage{algorithm}%
\usepackage{algorithmicx}%
\usepackage{algpseudocode}%
\usepackage{listings}%
\usepackage{caption}
\usepackage{subcaption}
\usepackage{fixmath}
\usepackage{float}
\usepackage{fancyhdr} 

\usepackage{hyperref}

\newcommand{\emailAGF}{agarciaflol@gmail.com}
\newcommand{\emailALC}{ale.lage@gmail.com}
\newcommand{\emailRM}{roberto.mulet@gmail.com}


\title{Looking into informal currency markets as Limit Order Books: impact of market makers}

\author[1]{%
    Alejandro García Figal%
}
\author[1,2]{%
    Alejandro Lage Castellanos%
}
\author[1,2]{%
    Roberto Mulet\textsuperscript{*}
} 

\affil[1]{%
    Group of Complex Systems and Statistical Physics, Faculty of Physics, University of Havana, La Habana, 10400, Cuba
}
\affil[2]{%
    Department of Theoretical Physics, Faculty of Physics, University of Havana, La Habana, 10400, Cuba
} 

\hypersetup{
pdftitle={Informal Currency Markets as Limit Order Books},
pdfsubject={q-fin.TR},
pdfkeywords={control theory, dynamic programming, limit order book, market maker},
}

\fancypagestyle{firstpage}{
    \fancyhf{} 
    \fancyfoot[C]{%
        \begin{minipage}{\textwidth}
            \centering
            \footnotesize
            Alejandro García Figal (\href{mailto:\emailAGF}{\emailAGF}) \\
            Alenadro Lage-Castellanos (\href{mailto:\emailRM}{\emailALC}) \\
            Roberto Mulet (\href{mailto:\emailRM}{\emailRM})
        \end{minipage}
    }
}

\begin{document}
\maketitle
\thispagestyle{firstpage}

\begin{abstract}
This study pioneers the application of the market microstructure framework to an informal financial market. By scraping data from websites and social media about the Cuban informal currency market, we model the dynamics of bid/ask intentions using a Limit Order Book (LOB). This approach enables us to study key characteristics such as liquidity, stability and volume profiles. We continue exploiting the Avellaneda-Stoikov model to explore the impact of introducing a Market Maker (MM) into this informal setting, assessing its influence on the market structure and the bid/ask dynamics. We show that the Market Maker improves the quality of the market. Beyond their academic significance, we believe that our findings are relevant for policymakers seeking to intervene informal markets with limited resources.
\end{abstract}

\section{Introduction}
\label{intro}

The rapid advancement of digital technology  and the availability of vast amounts of data capturing interactions, intentions, and actions of economic agents has facilitated a detailed examination of many financial ecosystems \cite{gueant2016financial, Bouchaud_Bonart_Donier_Gould_2018}. In this context, the study of market microstructure provides a theoretical framework to understand the dynamics of asset prices based on the analysis of trading systems, types of order and the interactions between the participants in the markets\cite{madhavan2000market}. It is safe to say that the scientific community has a reasonable understanding about the statistical properties of most financial markets\cite{gueant2016financial, Bouchaud_Bonart_Donier_Gould_2018, slanina2014econophysics}.

However, most of this comprehension reflects, at best, the reality of formal markets in developed countries\cite{reinganum1990market, muscarella2001market, dominguez2003market, venkataraman2007value, mizrach2014market}, and leaves a significant gap in our understanding of informal financial markets. These are more prevalent in the Global South where they help in providing access to financial services to individuals and small businesses, often excluded from formal banking systems. Understanding the microstructure of these markets can help policymakers design interventions that enhance financial inclusion and promote economic development.

This gap in our understanding results from the scientific bias favoring the research about structures in the developed world, but also from lack of confident data to quantify informal markets. The appearance of internet and social media, and the displacement of these markets there, creates an opportunity to alleviate this situation. This works take profits of this and focus its attention on the Cuban informal currency market\cite{garcia2024efficiency, vidal2024using}.

In the first part of the manuscript, we exploit real data of bid/asks intentions extracted from informal electronic exchange sites in Cuba and model the dynamics of these intentions as the one of a Limit Order Book (LOB) \cite{gould2013limit}. We hypothesize that this approach is a proxy for the dynamics of the real market, and therefore that it could elucidate some of its properties, including liquidity, stability, market volume profiles, and order distributions \cite{gould2013limit, bouchaud2002statistical, saddier2024bayesian, potters2003more}. In the second part we study the impact of a market maker in the structure of this market \cite{klock1999impact, venkataraman2007value, zhu2009does, anand2016market}.

To the best of our knowledge, this is the first application of the market microstructure framework to an informal financial market and the first time, that the performance of a market maker is estimated in a similar context.

The work is organized as follows: In Section \ref{lob} we present the model of the LOB. We continue with section \ref{properties} presenting and discussing the empirical statistics obtained.  We then introduce the market maker framework and discuss its impact on the structure of the market in Section \ref{market_maker}. We continue modeling the impact of the market marker on the bid/asks dynamics of the informal market in Section \ref{pricing}. Finally we present the conclusions of our work. All the data required to reproduce the results presented in this study are available in this \href{https://github.com/lolfig/Looking-into-Informal-Currency-Markets-as-Limit-Order-Books.git}{git-hub reppository}.

\section{The Limit Order Book}
\label{lob}

Most financial markets works through an electronic algorithm called, Limit Order Book (LOB) \cite{Bouchaud_Bonart_Donier_Gould_2018, bouchaud2002statistical, gould2013limit, coletta2022learning, gopikrishnan2000statistical}. It facilitates the interaction of buyers and sellers sorting the orders by price,  from most attractive to least attractive (highest to lowest for buy orders and lowest to highest for sell orders). The orders are executed giving priority to more atractive prices and if  multiple orders share the same price at the time of execution, priority is determined by arrival time.

We will  assume that the electronic orders present in social media platforms studied as LOB represent a proxy of the real informal currency market. The orders are obtained by web scraping the principal site of currency exchange in Cuba (ElTOQUE \cite{ElToque}) to gather public intentions to purchase or sell currencies. The scraping is conducted using a python-based system, which allows us to extract sales and purchases intentions from posts and comments in these web page. To date, we have collected messages from July 2021 to February 2025, resulting in a total of $814,233$ orders, representing $95.11\%$ of the total messages collected. The remaining messages were "unclear" and lacked sufficient information for automatic classification. The messages are processed using regular expressions to identify patterns\cite{garcia2024efficiency}. We validated this automated process against human interpretation on a sample of messages.

We focus our attention here on the USD, the most demanded currency in the Cuban market, which also serves as a reference for other transactions. We process the messages to extract declared orders for buying or selling specific USD volumes at prices no worse than stated. Each order $x$ is characterized by four attributes: the sign $\epsilon_x$, $+1$ for a buy order and $-1$ for a sell order, the price $p_x$ (in cuban pesos, CUP), the volume $v_x > 0$, (defined in dollars, USD) and a timestamp $t_x$. Thus, an order $x$ can be defined as $x := (\epsilon_x, p_x, v_x, t_x)$. 

\begin{figure}[!ht]
    \centering
    \includegraphics[width=0.5\linewidth]{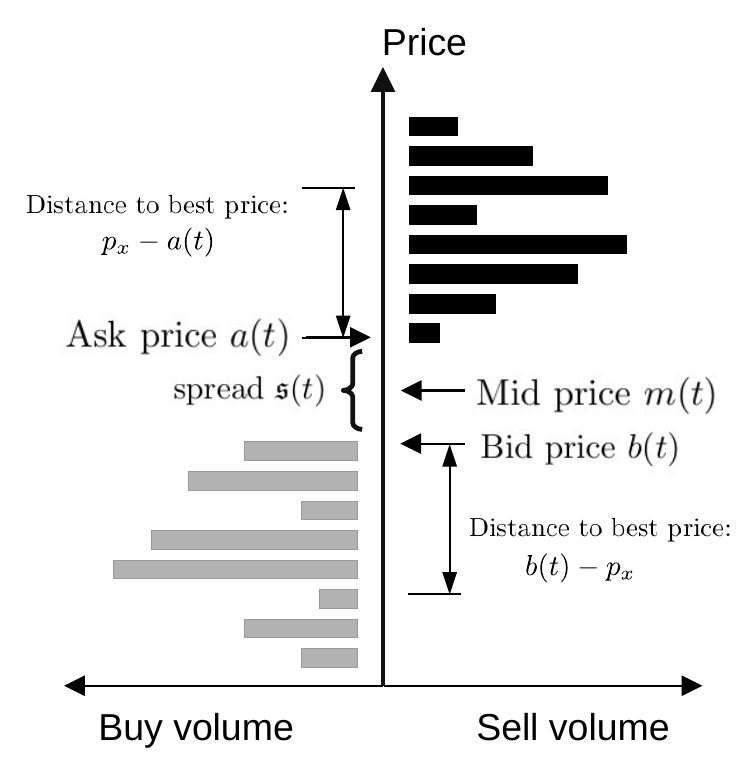}
    \caption{A schematic of an LOB, to illustrate the bid-price, the ask-price, the available volumes, the mid-price, and the bid-ask spread.}
    \label{fig:lob}
\end{figure}

When a buy order (or a sell order) $x$ comes in, the matching algorithm checks whether it is possible for the order $x$ to match a sell order (or a buy order) $y$, i.e., they match if $p_y \leq p_x$ (or $p_y \geq p_x$). If a match is possible, a transaction occurs immediately at the agreed quantity and price. Any portion of $x$ that is not executed instantly becomes part of the LOB at price $p_x$, and remains on the book until it is executed by an incoming order or canceled after seven days\footnote{In informal markets, where we proxy orders from raw messages, we assume an order is canceled after seven days.}.  Thus, an LOB can be defined as a collection $\mathfrak{L}(t)$ of unfulfilled intentions to buy or sell an asset at a given time $t$. It can be  partitioned into a set of buy limit orders $\mathfrak{B}(t)$ with $\epsilon = +1$ and a set of sell limit orders $\mathfrak{A}(t)$ with $\epsilon = -1$ (Figure \ref{fig:lob}). The bid and ask prices are defined as:
\[ b(t) := \max_{x \in \mathfrak{B}(t)} p_x \quad \mbox{and} \quad a(t) := \min_{x \in \mathfrak{A}(t)} p_x\]
respectively. By following this dynamic, we can extract intraday information that is valuable, for example: the mid price  $m(t) := \frac{1}{2} [a(t) + b(t)]$, the bid-ask spread, $\mathfrak{s}(t) := a(t) - b(t) > 0$ and the distance to the best price:

\begin{equation*}
d(p_x, t) = 
    \begin{cases} 
        b(t) - p_x & \text{for } \epsilon_x=+1 \\ 
        a_x - p_x & \text{for } \epsilon_x=-1 
    \end{cases}
\end{equation*}
that are necessary for quantifying market liquidity and stability. We interpret the data processed as a comprehensive record of microscopic behavior of the market defining also the temporal evolution of visible liquidity and the interplay between order flow transactions and price changes.

In the next section we present statistical results describing the price behavior, spread distribution, confidence, and order flow of this market. We will demonstrate that these statistical properties exhibit distinct patterns, different from those observed in ``formal" financial markets.

\section{Statistical properties of the Limit Order Book}
\label{properties}

\begin{figure}[!ht]
    \centering
    \begin{subfigure}[b]{0.9\linewidth}
        \centering
        \includegraphics[width=\linewidth]{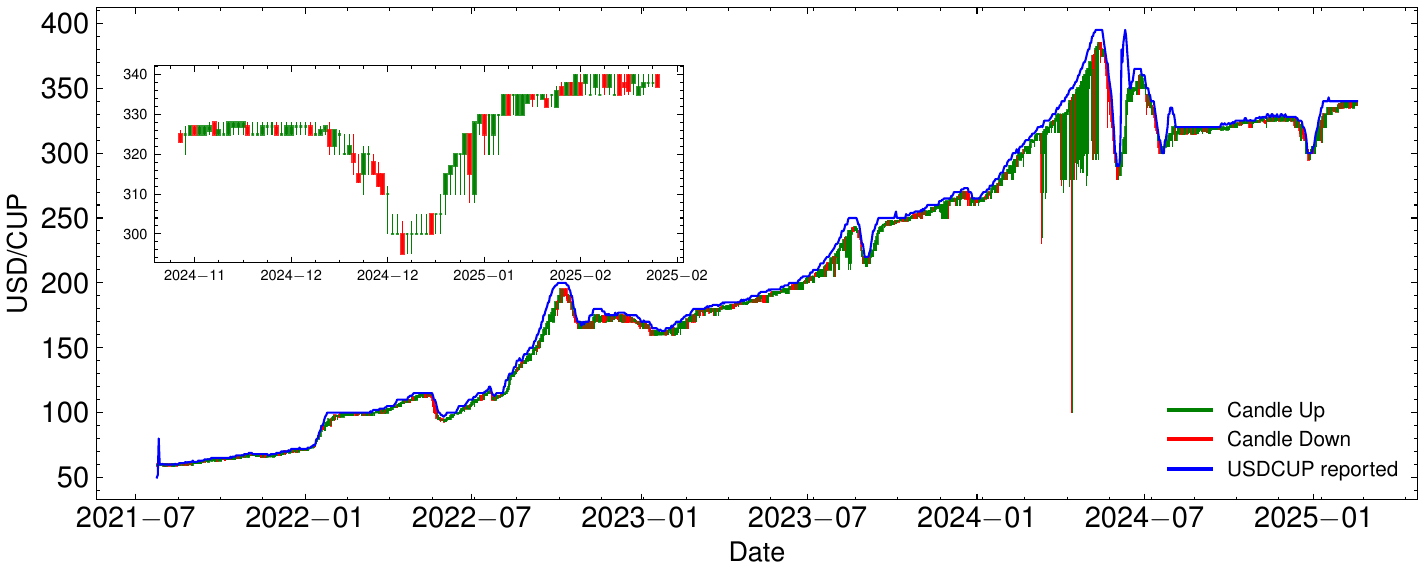}
        \caption{}
        \label{fig:candle}
    \end{subfigure}
    \hfill

    \begin{subfigure}[b]{0.9\linewidth}
        \centering
        \includegraphics[width=\linewidth]{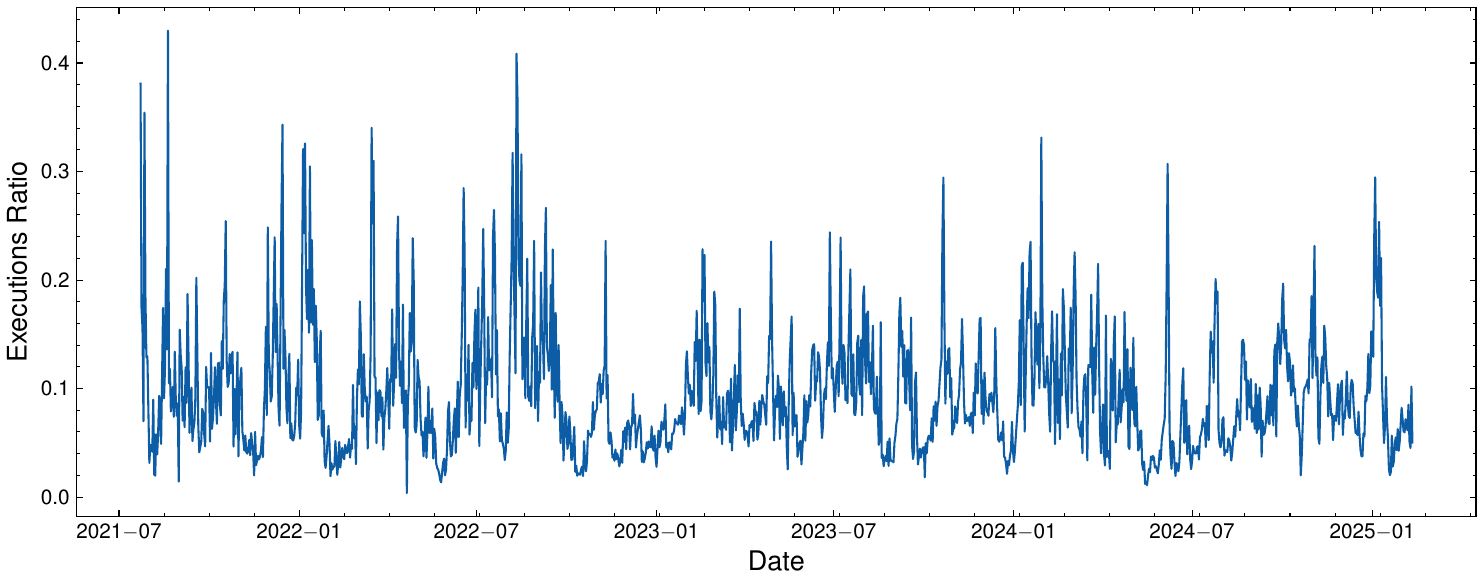}
        \caption{}
        \label{fig:executions}
    \end{subfigure}\
        \caption{a) Japanese candlestick chart illustrating the dynamics of the USD strike price in this market, the blue time series is the USD price reported on the elTOQUE website. b) Data on executions that occurred in response to the dynamics of the limit order book (LOB), displaying the execution ratio relative to the total volume for the day.}
    \label{fig:market_dynamics}
\end{figure}

To illustrate the fluctuations in the USD price in Figure \ref{fig:candle} we present the simulated dynamics of a  LOB on this informal market using Japanese candlestick chart. A green candlestick indicates a bullish day (when the closing price is higher than the opening price), while a red candlestick signifies a bearish day. It is important to note that the overall price dynamics do not differ significantly from that presented on the main reference website elTOQUE \cite{ElToque} (blue time series), or our previous estimations based on the Walrasian auction (marginal price) between the messages \cite{garcia2024efficiency}.

In Figure \ref{fig:executions} we show the daily ratio between the volume executed and the total volume. The results show a very low rate of execution, never exceeding \(40\%\) and averaging around \(10\%\). In comparison, stable and formal financial markets typically exhibit a much higher ratio, where the volume in the LOB is a small fraction of the corresponding daily traded volume \cite{Bouchaud_Bonart_Donier_Gould_2018}. This suggest that within this dynamics,  the majority of the intentions remain unfulfilled, casting suspicion on an illiquid and unstable market.

\begin{figure}[!ht]
    \centering
    \begin{subfigure}[b]{0.9\linewidth}
        \centering
        \includegraphics[width=\linewidth]{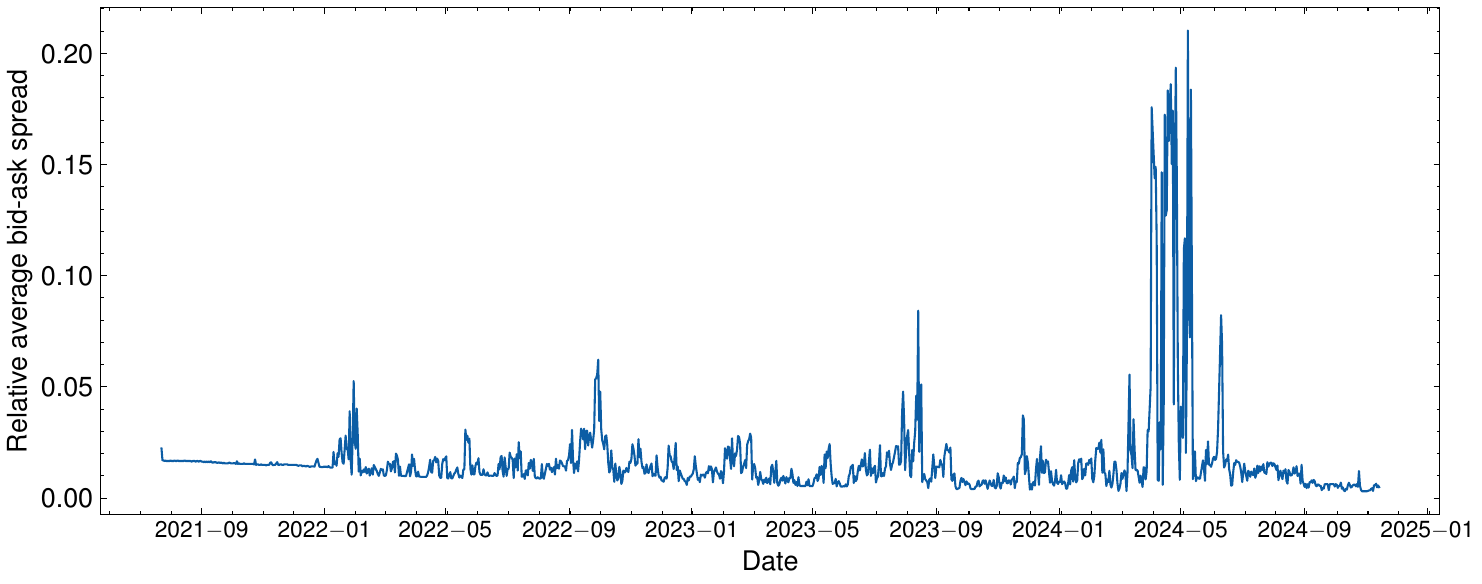}
        \caption{}
        \label{fig:spread}
    \end{subfigure}
    \hfill

    \begin{subfigure}[b]{0.9\linewidth}
        \centering
        \includegraphics[width=\linewidth]{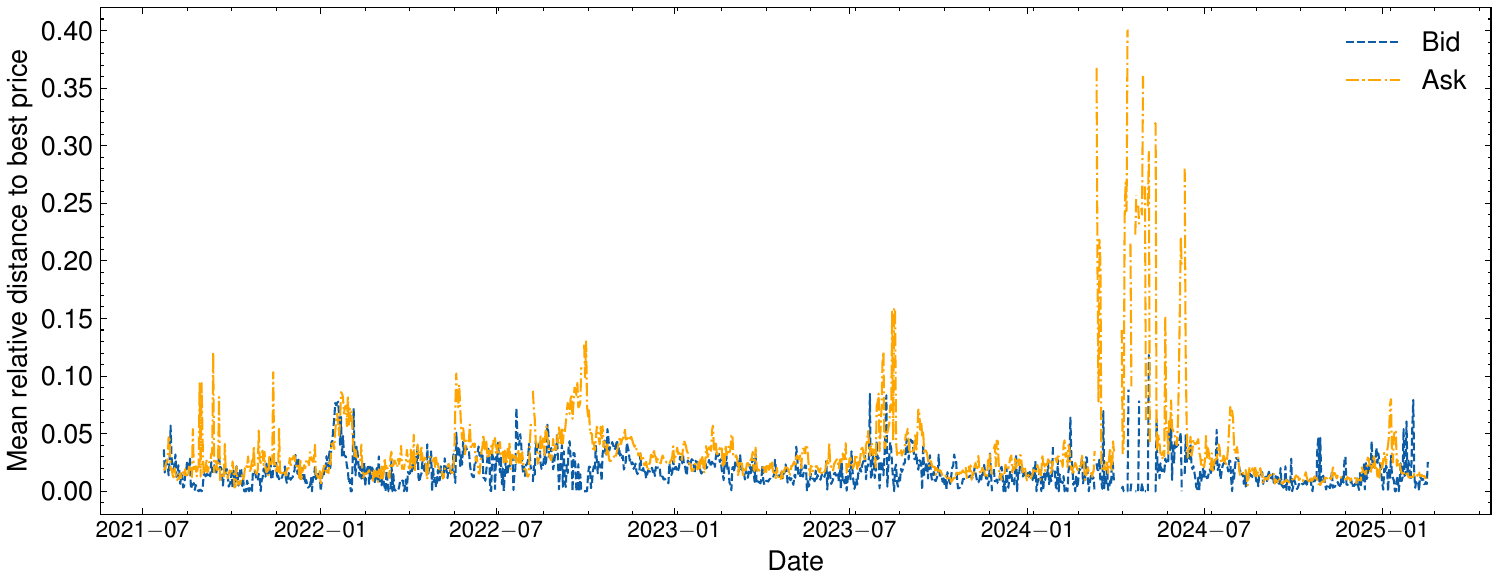}
        \caption{}
        \label{fig:distance}
    \end{subfigure}

    \caption{a) Relative average of the bid-ask spread shown at the top, with relative intraday spreads ($s/m$) calculated and averaged for each day. b) Daily mean relative distance to the best price, with best bid price (blue) and best ask price (orange).}
    \label{fig:market_dynamics2}
\end{figure}

This is also explored in  Figure \ref{fig:spread} where we plot the bid-ask  spreads of the market. A ``tight" spread indicates that there are many buyers and sellers in the market, which facilitates the execution of orders without significantly affecting the price of the asset (less costs for trades). In a stable, trustworthy and liquid environment, the spread is expected to be low, around  $10^{-2}\%$ and $10^{-3}\%$ relative to the mid-price \cite{Bouchaud_Bonart_Donier_Gould_2018}. On the contrary, in this informal market two patterns are observed: irregular and large spreads. These results demostrate that the market is illiquid. Agents frequently quote prices that are significantly distant from the mid-price, reflecting their lack of confidence in the reference price.

This lack of confidence is further illustrated in Figure \ref{fig:distance}, which shows that the average distance that agents quote from the best price remains relatively large compared to formal markets \cite{Bouchaud_Bonart_Donier_Gould_2018}. Notably, there are also periods marked by heightened uncertainty, indicated by peaks in this distance. But in general we should remark that independently of the clear positive drift of the USD price during the last four years (see Figure \ref{fig:candle}), if interpreted within a LOB scenario, the properties of the relative fluctuations of the market are stable.

\begin{figure}[!htb]
    \begin{minipage}{0.45\linewidth}
        \centering
        \includegraphics[width=\linewidth]{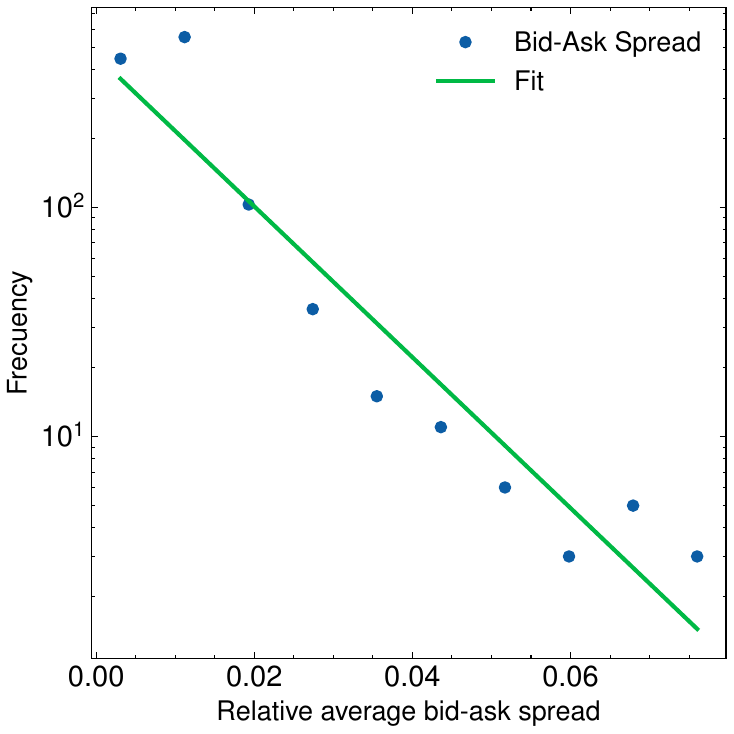}
    \end{minipage}
    \hfill
    \begin{minipage}{0.45\linewidth}
      \centering
      \includegraphics[width=\linewidth]{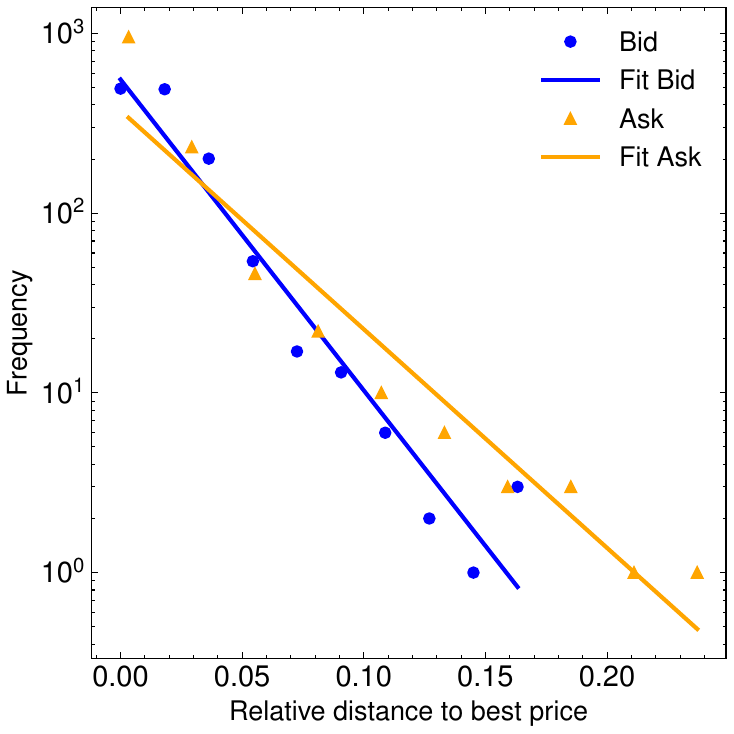}
    \end{minipage}
    \caption{Semi-logarithmic scale histograms of the relative average bid-ask spread in the left and the relative distance $d$ for buyers and sellers in the right. The standard behavior in both plots follows an exponential distribution.}
    \label{fig:spread_distance}
\end{figure}

To get a deeper insight in the left panel in Figure \ref{fig:spread_distance} we show the probability distribution of the daily bid-ask spread during the whole period. This distribution can be fitted with an exponential function indicating  that changes have a characteristic scale. This  suggests that as time progresses, the probability of observing larger deviations from the mean decreases exponentially, i.e., over time,  there is a higher probability that the series will revert to the mean. Moreover, also the relative average distance to the best bid and ask prices distributes exponentially (right panel Figure \ref{fig:spread_distance}). Buyers and sellers exhibit similar exponential patterns in their quoting behavior. While usually this distance is driven by market participants continuously monitoring $b(t)$ and $a(t)$ to guide their actions this may not hold true in this informal context, where this information is unavailable. Nonetheless, an exponential trend is clearly observed. 

\begin{figure}[!htb]
    \begin{minipage}{0.48\linewidth}
        \centering
        \includegraphics[width=\linewidth]{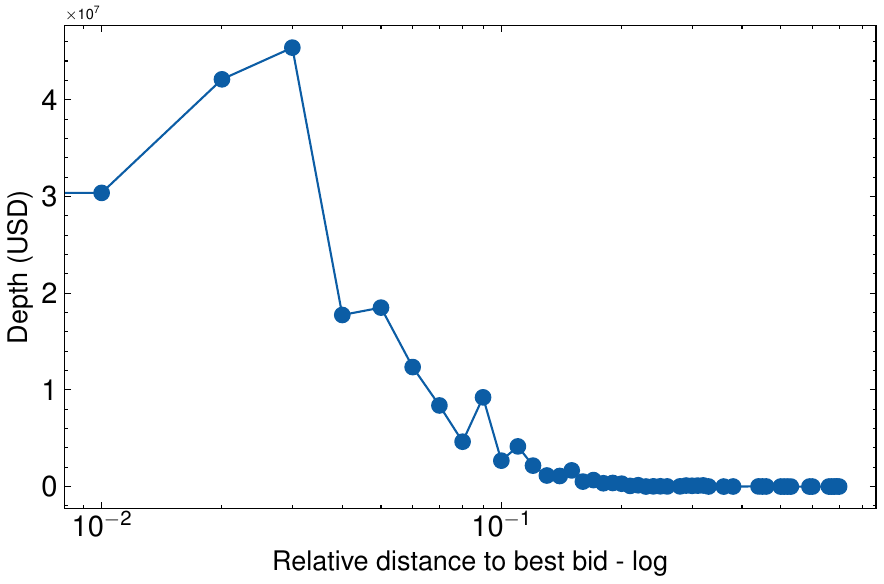}
    \end{minipage}
    \hfill
    \begin{minipage}{0.48\linewidth}
        \centering
        \includegraphics[width=\linewidth]{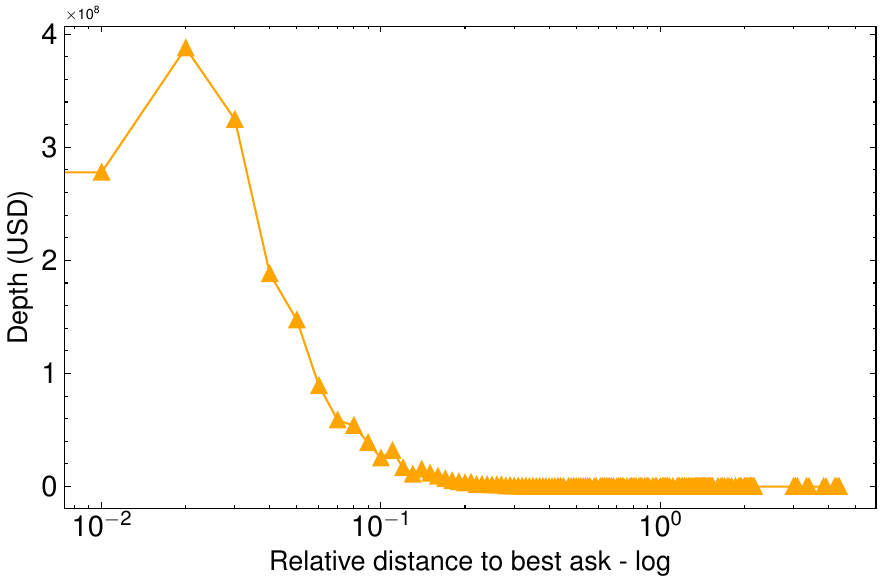}
    \end{minipage}
    \caption{Mean relative volume profiles, measured as a function of the distance from the best price. The x axis is in logarithmic scale. The left panel shows the bid size, while the right panel displays the adsk size.}
    \label{fig:volume_profile}
\end{figure}

Figure \ref{fig:volume_profile} illustrates the mean relative volume profiles as a function of the distance $d$ from the best price. The profile initially increases for small distances, reaching a maximum and then declines rapidly at larger distances. Such peaking and declining behavior is more common and pronounced in markets with a large tick size \cite{Bouchaud_Bonart_Donier_Gould_2018}. However, in these markets, the decline over larger distances is much slower. The pronounced decrease is another sign of the illiquidity in the market.

\begin{figure}[!htb]
    \centering
    \begin{minipage}{0.48\linewidth}
        \centering
        \includegraphics[width=\linewidth]{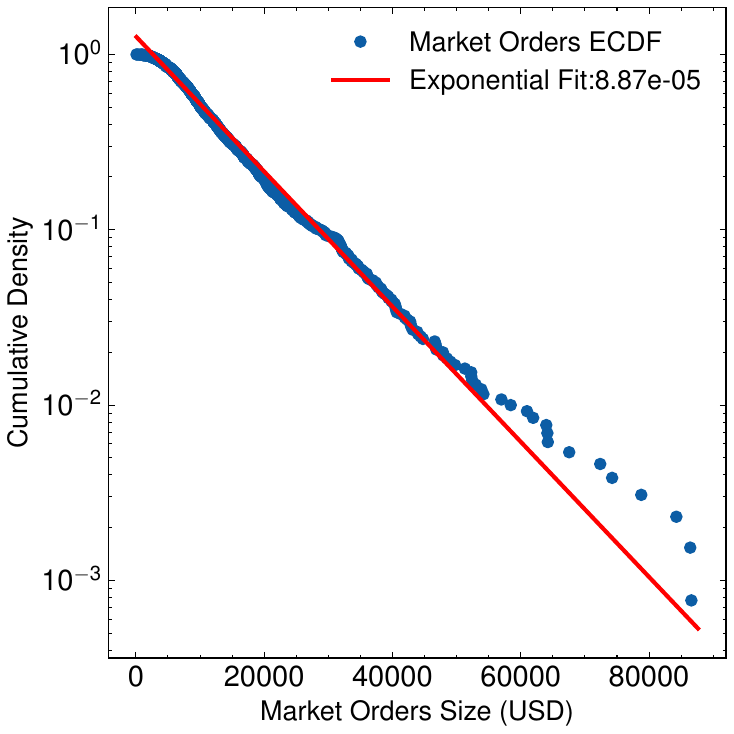}
    \end{minipage}
    \hfill
    \centering
    \begin{minipage}{0.48\linewidth}
        \centering
        \includegraphics[width=\linewidth]{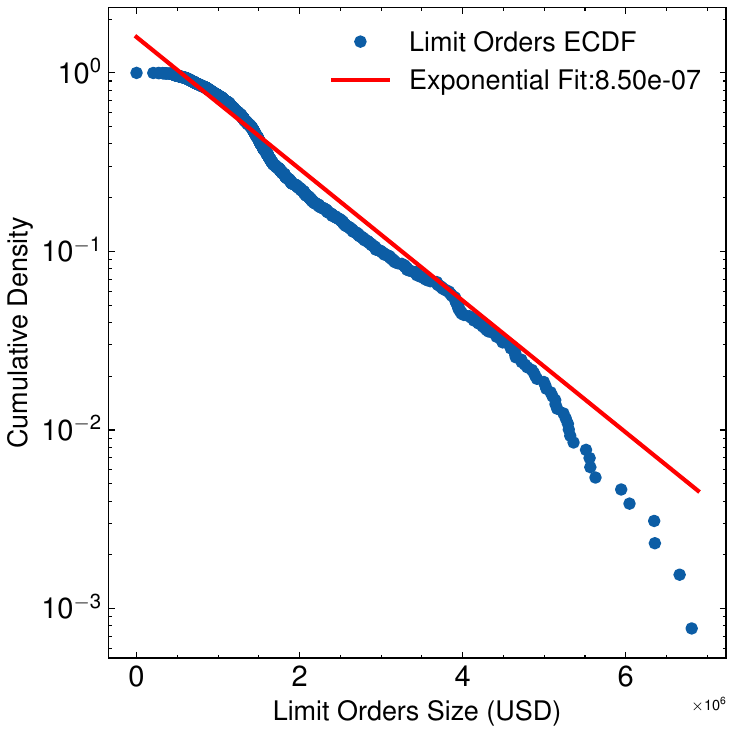}
    \end{minipage}
    \caption{Empirical cumulative density functions (ECDFs) of the sizes of market orders (left) and limit orders (right), expressed in terms of lots (USD). The plots are presented on a semi-logarithmic scale, allowing exponential distributions to appear as straight lines.}
    \label{fig:orders_size}
\end{figure}

Finally, the empirical cumulative density functions (ECDFs\footnote{The cumulative density function (CDF) is plotted as $1-CDF$ to highlight the tail behavior of the distribution, which is particularly useful for visualizing exponential decay on a logarithmic scale.}) of order sizes for limit and market orders is illustrated in Figure \ref{fig:orders_size}, presented in lots (USD) on a semi-logarithmic scale. While the most common order sizes are relatively small, typically a few thousand dollars, there are also significantly larger orders present. Specifically, the distribution of market orders exhibits an exponential decay. This contrasts with findings in several studies that report power law distributions \cite{Bouchaud_Bonart_Donier_Gould_2018, gopikrishnan2000statistical, maslov2001price, gabaix2003theory}, indicating that order sizes do not cluster around a typical average. In this informal market, however, order sizes change at a rate proportional to their current value, suggesting an environment where large orders are  rare and have a small impact on the market dynamics.

In summary, our analysis elucidates the statistical properties of the informal market interpreted as a Limit Order Book. Our results  reveals an informal market characterized by illiquidity, a lack of confidence, and an exponential distribution of spreads and order sizes. Moreover, as we show below, it also provides valuable insights for simulating and optimizing the behavior of a Market Maker. For instance, the empirical exponential decay observed in the distribution of market order sizes (Figure \ref{fig:orders_size}) offers critical information for estimating the intensity of order arrivals, which is essential for calibrating the market maker's pricing strategy based on buying and selling prices. 

\section{Optimal pricing: the market maker}\label{market_maker}

Operating in a limit order book, a market maker is a liquidity provider who continuously quotes prices at which they are willing to buy and sell an asset \cite{ho1981optimal, avellaneda2008high}. This type of agent is prevalent in nearly all markets and plays a crucial role in the price formation process. There are various types of market makers, {\em officials} like designated market makers (DMMs) have formal agreements with exchanges to maintain fair and orderly markets \cite{venkataraman2007value}. Others, such as certain high-frequency traders, act as liquidity providers without any obligations \cite{avellaneda2008high}. It is empirically known that some market makers exist in the Cuban informal market, but they operate with simple rules setting prices above and below the reference public price for the USD. In this work, we simulate the effect of the latter in this type of market and test the profit margin of such optimal type of traders. For simplicity we consider that only one market maker operates within the entire market. Moreover, this agent exploits the well-known strategy proposed by Avellaneda-Stoikov in \cite{avellaneda2008high}.

A detailed analysis of the Avellaneda-Stoikov formalism is provided in \cite{ho1981optimal, avellaneda2008high, gueant2013dealing}. For completeness, we reproduce the key elements of this framework in Appendix \ref{A2}, while a brief overview of the Hamilton-Jacobi-Bellman equation is presented in Appendix \ref{A1}. In essence, the model assumes that the market maker operates in a single asset (e.g., USD) and that the mid-market price follows a diffusion process:

\begin{equation}
    ds_t = \sigma dW_t
\end{equation}
where $W_t$ represents the standard one-dimensional Brownian motion. The market maker at each instant in time proposes a purchase price $s^b_t = s_t - \delta^b_t$ and a sale price $s^a_t = s_t + \delta^a_t$. The execution probabilities depend on $\delta^b_t$ and $\delta^a_t$, which determine the distance from the mid-price. We model the number of assets bought and sold as Poisson processes:
\[ N^b_t \sim \text{Poisson} (\lambda^b_t dt) \quad \mbox{and}\quad  N^a_t \sim \text{Poisson} (\lambda^a_t dt)\]
with intensities defined in \cite{avellaneda2008high} as $\lambda(\delta) = \Lambda e^{-k \delta}$. This functional form reflects the empirical relationship between the distance $\delta$ from the mid-price and the rate at which orders are executed at that price.
\begin{figure}[!ht]
    \centering
    \includegraphics[width=0.8\linewidth]{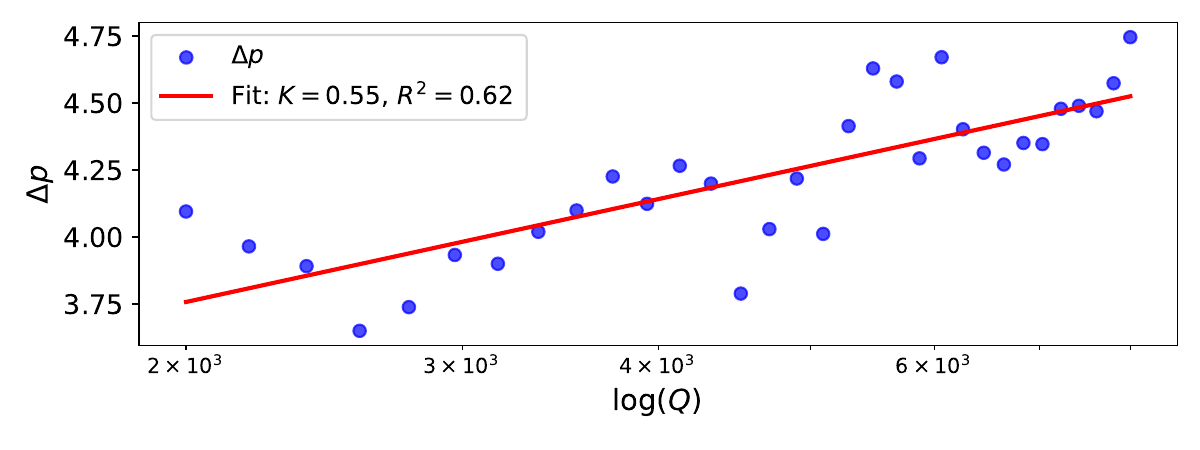}
    \caption{Price response function $\Delta p$ as a function of $\ln Q$. The periods corresponds to April 2023.}
    \label{fig:price_impact}
\end{figure}
The standard derivation of $\lambda(\delta)$ is based on two empirical observations. First, the size distribution of market orders follows a power law \cite{gopikrishnan2000statistical, maslov2001price, gabaix2003theory}:
\begin{equation*}
    Q(x) \propto x^{-1-\alpha}
\end{equation*}
Second, the price impact function $\Delta p = \left\langle \epsilon(t) [m(t+\tau)-m(t)]|Q \right\rangle$ which measures the change in the mid-price following a market order, exhibits a logarithmic dependence on $Q$ \cite{potters2003more}:
\begin{equation*}
    \Delta p \propto ln(Q)
\end{equation*}
This second relationship is also observed in the simulated market modeled as a LOB, as illustrated in Figure \ref{fig:price_impact}. However, in our informal market setting, we observe a different distribution for market order sizes (Figure \ref{fig:orders_size}). Specifically, we find that $Q(x) \propto e^{-\alpha x}$. Incorporating this information, we derive the following intensity function:
\begin{align}
    \lambda(\delta) &= \Lambda P(\Delta p > \delta) 
    \nonumber
    \\
    &= \Lambda P(\ln(Q) > K \delta) 
    \nonumber
    \\
    &= \Lambda P(Q > e^{K \delta}) 
    \nonumber
    \\
    &= \Lambda \int_{e^{K \delta}}^{\infty} e^{-\alpha x} \, dx 
    \nonumber
    \\
    &= \frac{\Lambda}{\alpha} e^{-\alpha e^{K \delta}} \label{eq:resultado}
\end{align}
where $\Lambda$ represents the frequency of market buy or sell orders, estimated by dividing the total daily traded volume by the average size of market orders, and $K$ is the slope of the fitted line in the equation $\Delta p = K ln(Q) + \varepsilon$.

In our simulations we estimated $\alpha$ from the actual empirical data of order frequency and size distribution from the market orders (Figure \ref{fig:orders_size}). The inventory is modeled by the process $q_t= N^b_t - N^a_t$ and inventory wealth (the cash) also becomes stochastic as it changes with each transaction, $dc_t = p^a dN^a_t - p^b dN^b_t$. Finally, the market maker aims to maximize the expected value of their CARA utility function $u(s,c,q,s) = -e^{-\gamma(c_t + q_t s_t)}$:
\begin{equation}
    v(s,c,q,t) = \max_{\delta^a, \delta^b} E_t[-e^{-\gamma(c_T + q_T S_T)}] \label{eq:valuefunctionstoikovs}
\end{equation}
Note that the control parameters are the deltas which indirectly influence orders received by the market maker. Using the principles of Dynamic Programming and one can see that the value function $v$ solves the following Hamilton-Jacobi-Bellman (HJB) equation \cite{ho1981optimal,avellaneda2008high}:
\begin{align}
    \frac{\partial v}{\partial t} + \frac{1}{2}\sigma^2 \frac{\partial^2 v}{\partial s^2} &+ \max_{\delta^b}[\lambda^b(\delta^b)(v(s,c-s+\delta^b,q+1,t) - v(s,c,q,t))]
    \nonumber
    \\
    &+\max_{\delta^a}[\lambda^a(\delta^a)(v(s,c+s+\delta^a,q-1,t) - v(s,c,q,t))] = 0, 
    \nonumber
    \\
    &v(s,c,q,T)=-\exp{(-\gamma(c+qs))}
\end{align}
However, through a series of educated ansatz, Avellaneda and Stoikov showed that the distances $\delta^a$ and $\delta^b$ can be computed by a simpler relation expressesing this spread around the indifference price $r(s,q,t)=s-q \gamma \sigma^2(T-t)$ for the market maker:
\begin{align}
    \delta^b &= \gamma q \sigma^2 (T-t) + \frac{2}{\gamma} \ln(1+\frac{\gamma}{k}) \\
    \delta^a &= -\gamma q \sigma^2 (T-t) + \frac{2}{\gamma} \ln(1+\frac{\gamma}{k})
\end{align}
We are employing the same procedure (see Appendix \ref{A3}), with the only difference being that our intensity function is the described in \ref{eq:resultado}. Using this approach, we obtain the following transcendental equations for the distances $\delta$:
\begin{align}
    \delta^b &= \gamma q \sigma^2(T-t) + \frac{1}{\gamma}\ln(1+\frac{\gamma}{aKe^{K\delta^b}})
    \label{transcdeltab}
    \\
    \delta^a &= -\gamma q \sigma^2(T-t) + \frac{1}{\gamma}\ln(1+\frac{\gamma}{aKe^{K\delta^a}}) \label{transcdeltaa}
\end{align}

This solution defines the bid and ask prices used by the market maker in this informal market. It is important to note that this model assumes a constant $\sigma$ in the dynamics of the mid-price. However, as demonstrated in \cite{garcia2024efficiency}, the price in the cuban currency market can be modeled as a hidden markov model with two states. To facilitate comparison, we conduct two kinds of simulations: one that estimates a single constant $\sigma$ from the time series, and another that updates $\sigma$ every 10 days. In both cases the initial conditions for the simulation are: cash at the initial time $c_0 = 1 \times 10^5$ CUP, initial inventory $q_0=0$, the risk factor $\lambda=0.1$, the slope $K=0.55$, $\alpha=-8.87e-05$ and $\sigma=2.38$.

We refer to this process as simulations because we use a bootstrap technique to artifially enlarge the dataset. In short, to test similar yet different scenarios, we generate sub-samples from the actual set orders received each day by randomly sampling with replacement. This creates a fresh set of daily orders, and consequently, a "new" state of the LOB. Although these states exhibit similar behavior, they are not identical. This process is iterated $n$ times, and we compare the outputs.

\begin{figure}[!ht]
    \centering
    \includegraphics[width=0.6\linewidth]{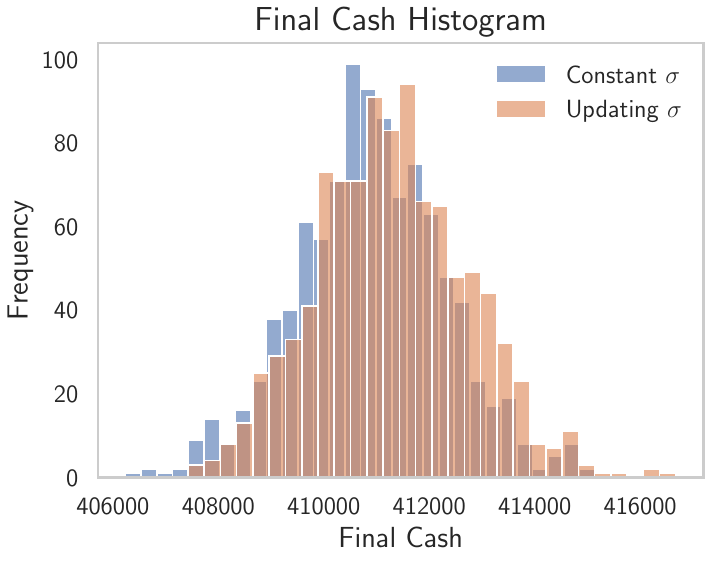}
    \caption{Histogram of final cash flows after 1000 simulations comparing the two strategies, using a fixed initial estimate of $\sigma$ and updating $\sigma$ every 10 days.}
    \label{fig:final_cash_hist}
\end{figure}

The results are summarized in Figure \ref{fig:final_cash_hist}. As shown, the final cash distributions under the two strategies are nearly identical, indicating that updating $\sigma$ every 10 days does not significantly alter the outcomes. Nevertheless, the optimization strategy increases the market maker's initial cash by a factor of four. While this result may seem unsurprising given the low execution rates and informal nature of the market, it underscores the potential profitability of market-making strategies in such settings.

\begin{figure}[!ht]
    \centering
    \includegraphics[width=0.8\linewidth]{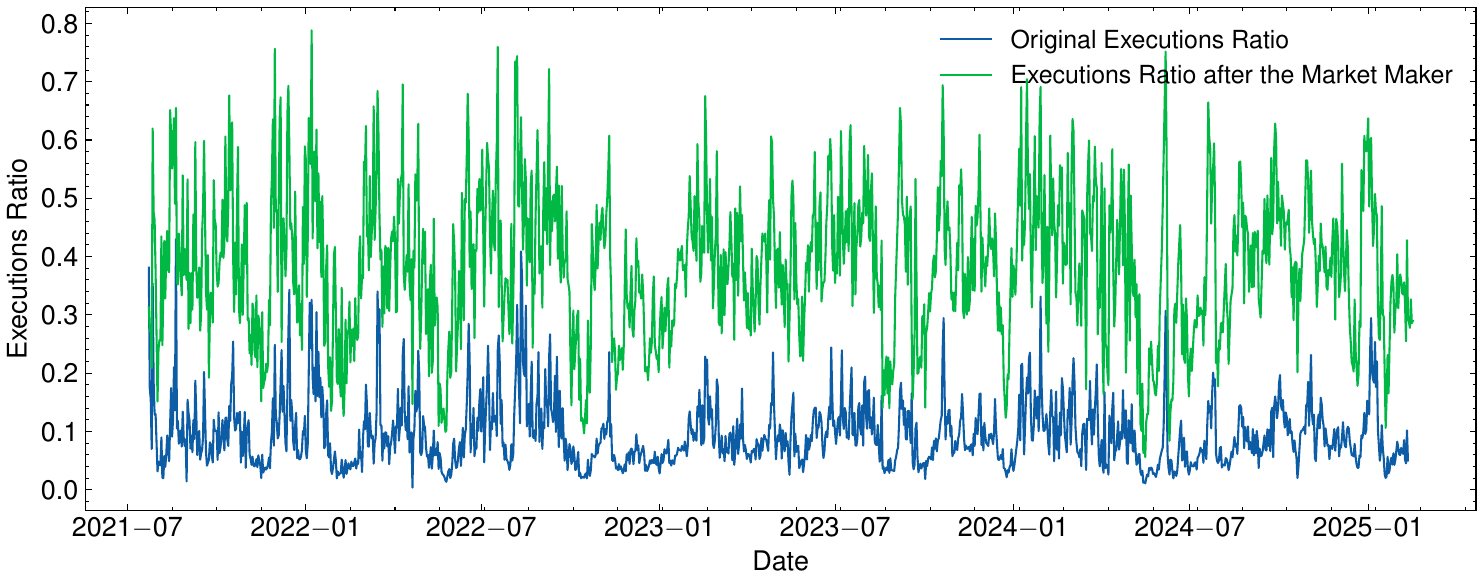}
    \includegraphics[width=0.8\linewidth]{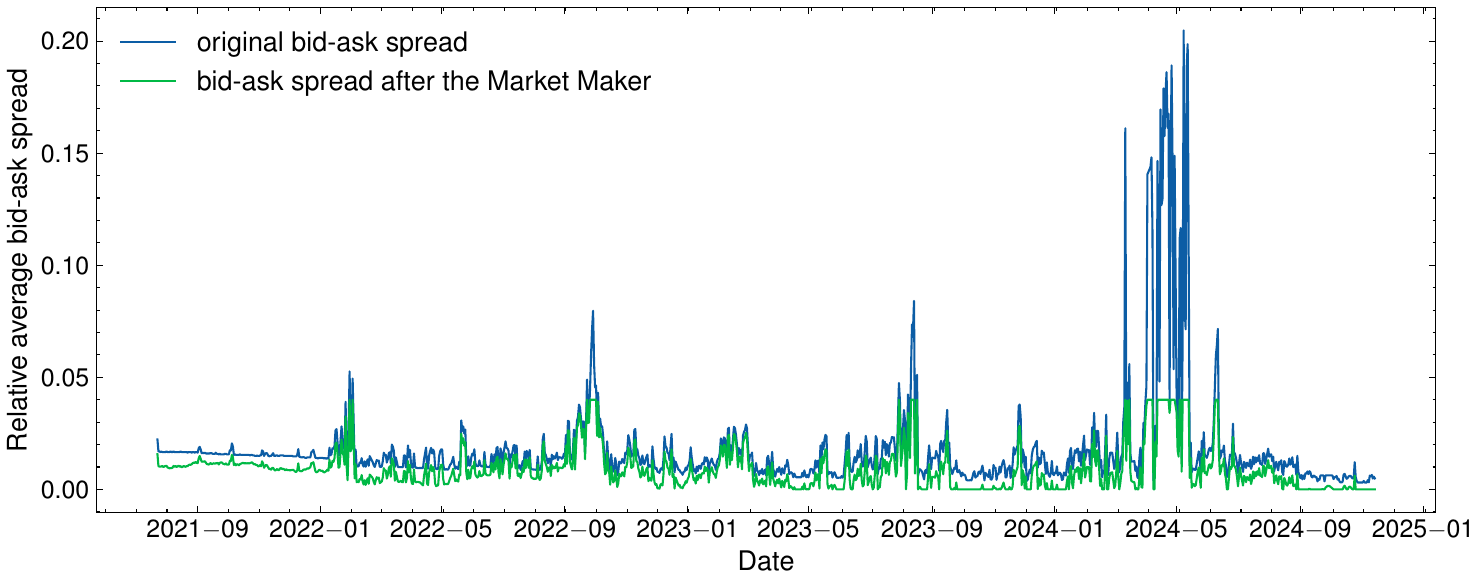}
    \caption{Simulation results of a market maker operating within the dynamics of a limit order book (LOB) in the informal market. The top panel presents the execution ratio: the blue line represents the original results following the LOB dynamics, while the green line incorporates the effects of a market maker. The bottom panel depicts the bid-ask.}
    \label{fig:mm_results}
\end{figure}

While we have successfully developed a market maker capable of generating substantial profits, we want to highlight the results presented in Figure \ref{fig:mm_results}. The introduction of the market maker led to significant improvements in both the bid-ask spread and the percentage of executions. Specifically, the number of executions increases to an average of $40\%$, while the spread decreases significantly, particularly in areas where liquidity was previously limited. Although these enhancements do not yet match the levels observed in formal markets, they demonstrate that the presence of a market maker influences positively the indicators of the market. This finding is particularly noteworthy and shows, for the first time, how a robust liquidity provider improves the properties of an informal market.

\section{Impact of Market Maker Introduction on Pricing}\label{pricing}

In this section we estimate the impact of introducing a market maker on the pricing dynamics\cite{potters2003more, zhu2009does, das2008effects}. Our approach is grounded in the following reasoning: Our original data set is a series of ex-ante orders (messages), each one with its own price. However, it is reasonable to expect that the presence of the market maker impacts the dynamics of price formation itself. Our intuition is that the mid-price based on the volume of USD in the LOB, is defined through the following Markov chain model
\begin{equation}
    \hat{s}_{t+1} = \hat{s}_t + \xi_t(Q^b_t - Q^a_t) + \epsilon_t
\end{equation}
where $\hat{s}_0 = s$, and $Q^b_t$ and $Q^a_t$ represent the amounts of USD in the LOB on the bid and ask sides respectively. The parameter $\xi_t$ captures the change in price resulting from imbalances between buy and sell volumes in the order book (point elasticity), defined as follows:
\begin{equation}
    \xi_t = \frac{\partial s_t}{\partial (Q^b_t - Q^a_t)}
\end{equation}
and in what follows we consider that although the imbalance $Q^b_t - Q^a_t$ changes in the presence of the market maker, $\xi$ the elasticity of the market, which is estimated from the actual data, does not. The intuition is that it reflects only an intrinisic sensitivity of the players. Then, each day, as limit orders fluctuate due to market maker activity, we derive a new $\hat{s}_t$ that diverges from $s_t$. The adjusted price for each order is then calculated using:
\begin{equation}
    \hat{p}_x = \frac{p_x \cdot \hat{s}}{s}
\end{equation}
where $p_x$ is the original price of the order. In essence, we create an artificial mid price series based on the observed elasticity. We then use the factor between the observed and predicted mid prices to scale each order price. By definition, this factor will be one if the total amount of LOB is not changed, and therefore is selfconsistent with the observed time series. When the market maker is present, and changes the LOB amounts, this methodology allows us to derive an updated USD price that reflects the influence of the market maker's activities. The pseudo-code illustrating the price update process can be found in Appendix \ref{A4}.

\begin{figure}[!ht]
    \centering
    \includegraphics[width=\linewidth]{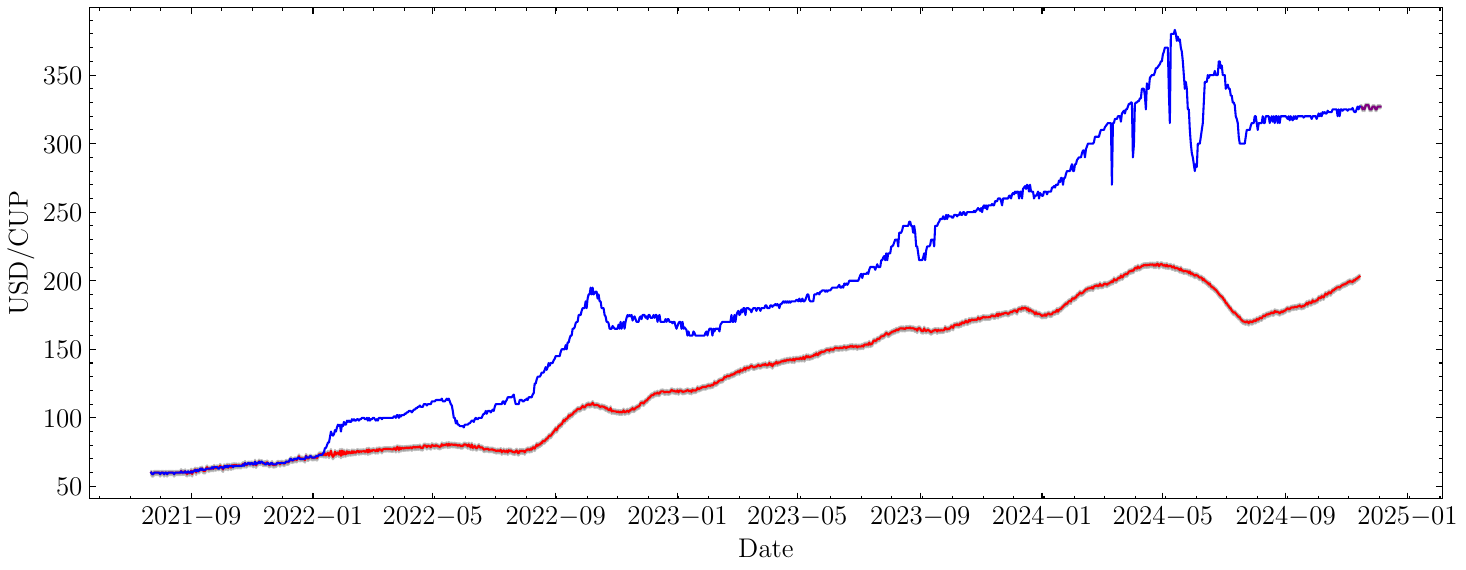}
    \caption{Time series of the USD/CUP exchange rate, showing updated prices derived from the implementation of the market maker strategy and the algorithm presented in Appendix \ref{A4}. The red time series represents the adjusted price ($\hat{s}$) when the market maker is introduced on the first day. The purple time series corresponds to the adjusted price (for 15 days) when the market maker is introduced on the last available day, with the new bid/ask dynamics for subsequent days simulated using the bootstrap technique.}
    \label{fig:prices_mm}
\end{figure}

The results of this framework are illustrated in Figure \ref{fig:prices_mm}, which shows the time series of USD/CUP exchange rates after applying the market maker strategy. Notably, the adjusted prices ($\hat{s}_t$) consistently remain below the original prices at the onset of the market maker's activity. This suggests that the market maker effectively reduces the mid-price by addressing volume imbalances, thereby ``stabilizing" the pricing dynamic. 

Figure \ref{fig:prices_mm} also highlights how the timing of the market maker's introduction influences pricing outcomes. Contrary to expectations of an immediate divergence, the adjusted prices initially appear to follow the prevailing trend set by the original price trajectory, regardless of when the market maker is introduced. This continuity suggests that the market maker's initial activity integrates seamlessly with the existing market dynamics, causing minimal disruption. However, over time, the influence of the market maker becomes more apparent, with the adjusted prices gradually diverging from the original prices. This delayed divergence may reflect the cumulative effect of the market maker's rebalancing activities, which require sufficient time to correct order imbalances and influence the overall pricing dynamics meaningfully. These observations underscore the importance of evaluating market maker strategies over an extended period to capture their full impact on price stabilization and trend modification in the LOB.

\section{Conclusions}\label{conclusions}

By employing a Limit Order Book (LOB) framework to model the dynamics of the Cuban informal currency market, we have identified several distinctive features that set it apart from formal financial markets. These include pronounced illiquidity, low execution rates, and the prevalence of large, irregular bid-ask spreads. Such characteristics underscore the unique challenges and inefficiencies inherent in informal markets, which contrast sharply with the relative stability and liquidity observed in formal markets, particularly those in developed economies.

A central contribution of this study lies in the exploration of a market maker’s role within this informal context. Through simulations, we demonstrate that the introduction of a market maker can markedly enhance market liquidity and stability. Specifically, the market maker reduces bid-ask spreads and increases execution rates, thereby improving overall market efficiency. While these improvements do not yet match the levels seen in formal markets, they suggest that market makers can serve as a stabilizing force in informal financial ecosystems, addressing some of their inherent inefficiencies.

Additionally, our analysis of price dynamics reveals that the market maker’s interventions lead to a gradual stabilization of the USD/CUP exchange rate. Over time, the adjusted prices diverge from the original prices, reflecting the market maker’s growing influence in addressing volume imbalances and correcting pricing inefficiencies. This finding highlights the By employing a Limit Order Book (LOB) framework to model the dynamics of the Cuban informal currency market, we have identified several distinctive features that set it apart from formal financial markets. These include pronounced illiquidity, low execution rates, and the prevalence of large, irregular bid-ask spreads. Such characteristics underscore the unique challenges and inefficiencies inherent in informal markets, which contrast sharply with the relative stability and liquidity observed in formal markets, particularly those in developed economies.

A central contribution of this study lies in the exploration of a market maker’s role within this informal context. Through simulations, we demonstrate that the introduction of a market maker can markedly enhance market liquidity and stability. Specifically, the market maker reduces bid-ask spreads and increases execution rates, thereby improving overall market efficiency. While these improvements do not yet match the levels seen in formal markets, they suggest that market makers can serve as a stabilizing force in informal financial ecosystems, addressing some of their inherent inefficiencies.

Additionally, our analysis of price dynamics reveals that the market maker’s interventions lead to a gradual stabilization of the USD/CUP exchange rate. Over time, the adjusted prices diverge from the original prices, reflecting the market maker’s growing influence in addressing volume imbalances and correcting pricing inefficiencies. This finding highlights the potential of market makers to mitigate short-term volatility and foster a more stable pricing environment in informal markets.

We think that although based on data from the Cuban informal currency market, our methodology can be extended to other markets and may offer valuable insights for policymakers and financial regulators in regions where informal markets play a significant role. By leveraging an understanding the market microstructure and the potential benefits of market makers, policymakers can design targeted interventions to enhance financial inclusion, promote economic development, and improve the stability of informal financial systems.potential of market makers to mitigate short-term volatility and foster a more stable pricing environment in informal markets.

We think that although based on data from the Cuban informal currency market, our methodology can be extended to other markets and may offer valuable insights for policymakers and financial regulators in regions where informal markets play a significant role. By leveraging an understanding the market microstructure and the potential benefits of market makers, policymakers can design targeted interventions to enhance financial inclusion, promote economic development, and improve the stability of informal financial systems.

\begin{appendices}

    \section{Hamilton-Jacobi-Bellman equation}\label{A1}
    
    The Hamilton-Jacobi-Bellman (HJB) equation is a partial differential equation that arises in the context of dynamic programming and optimal control theory.
    
    The approach here is that we want to minimize a cost function:
    \begin{equation*}
        J = \int_{t_0}^{t_f}L(x,u,t)dt
    \end{equation*}
    
    and this cost function si subject to a dynamic process:
    
    \begin{equation*}
        \dot{x}=f(x,u,t); x(0)=x_0 : \text{Fixed}
    \end{equation*}
    where $u$ is our control function.
    
    Now, for derivate the hamilton-jacobi-bellman equation, let us assume that we have obtained the optimal control $u^*$ (in the sate feedback form), and hence, the corresponding optimal trajectory $x^*$. Then the cost function is a function of the initial state $x^*_0=x_0$ and initial time $t_0$ ($t_f$ is fixed). The optimal cost function, also know as the value function ($V$), will be:
    \begin{equation*}
        V(t_0,x_0)=\int_{t_0}^{t_f}L(x^*,u^*,t)dt
    \end{equation*}
    Consider know a small time step $\Delta t$ between $t_0$ and $t_f$, so, we can write $V$ as:
    \begin{equation}
        V(t_0,x_0)=\int_{t_0}^{t_0+\Delta t}L(x^*,u^*,t)dt+\int_{t_0+\Delta t}^{t_f}L(x^*,u^*,t)dt \label{eq:V}
    \end{equation}
    Define:
    \begin{equation}
        J_0(x^*,u^*)=\int_{t_0}^{t_0+\Delta t}L(x^*,u^*,t)dt \label{eq:J}
    \end{equation}
    And by fundamental theorem\footnote{any part of an optimal trajectory is an optimal trajectory}:
    \begin{equation}
        \int_{t_0+\Delta t}^{t_f}L(x^*,u^*,t)dt=V((t_0+\Delta t), x_0(t_0+\Delta t)) \label{eq:L}
    \end{equation}
    Substituting \ref{eq:J} and \ref{eq:L} in \ref{eq:V} we get:
    \begin{equation}
        V(t_0,x_0)=J_0(x^*,u^*)+V((t_0+\Delta t), x_0(t_0+\Delta t)) \label{eq:vt_0,x_0}
    \end{equation}
    Now, assuming $L$ to be smooth over the interval $t_0$ to $(t_0+\Delta t)$ and $\Delta t$ to be "small enough", then:
    \begin{equation}
        J_0(x^*,u^*) = \Delta t L[x^*(t_0+\Delta t),u^*(t_0+\Delta t), (t_0+\Delta t)] \label{eq:j_0(x*,u*)}
    \end{equation}
    Next, we do Taylor's expansion in \ref{eq:L}:
    \begin{align}
        V((t_0+\Delta t), x_0(t_0+\Delta t)) &= V(t_0,x_0) + \frac{\partial V(t_0,x_0)}{\partial t_0 \partial x_0} + \dots 
        \nonumber
        \\
        V((t_0+\Delta t), x_0(t_0+\Delta t)) &= V(t_0,x_0) + \frac{\partial V(t_0,x_0)}{\partial t_0} \Delta t + \frac{\partial V(t_0,x_0)}{\partial x_0} \dot{x_0}\Delta t + \dots \label{eq:v_taylorexpansion}
    \end{align}
    where $\dot{x_0}\Delta t = \Delta x_0$.
    
    Combining \ref{eq:j_0(x*,u*)} and \ref{eq:v_taylorexpansion} in \ref{eq:vt_0,x_0}, and observing that $\dot{x_0}=f(x_0,u_0,t_0)$, we get:
    \begin{align}
        V(t_0,x_0) &= \Delta t L[x^*(t_0+\Delta t),u^*(t_0+\Delta t), (t_0+\Delta t)] + V(t_0,x_0)
        \nonumber
        \\
        &+ \frac{\partial V(t_0,x_0)}{\partial t_0} \Delta t + \frac{\partial V(t_0,x_0)}{\partial x_0} f(x_0,u_0,t_0)\Delta t \label{eq:vtoxs}
    \end{align}
    Canceling the term $V(t_0,x_0)$ from both sides in \ref{eq:vtoxs} and taking the limit $\Delta t \rightarrow 0$ leads to:
    \begin{equation}
        \frac{\partial V(t_0,x_0)}{\partial t_0} + L[x^*(t_0),u^*(t_0), t_0] + \frac{\partial V(t_0,x_0)}{\partial x_0} f(x_0,u_0,t_0) = 0 \nonumber
    \end{equation}
    Next, define:
    
    \begin{equation}
        \lambda \overset{\Delta}{=} \frac{\partial V(t,x)}{\partial x} \nonumber
    \end{equation}
    and for convenience (since we want to obtain the solution for any initial condition), we drop the subscript "0". Then we can write:
    
    \begin{equation}
        \frac{\partial V(t,x)}{\partial t} + L(x, u, t) + \lambda f(x,u,t) = 0 \nonumber
    \end{equation}
    Now, define $H_{opt} \overset{\Delta}{=} $ Hamiltonian for the optimum control. With this, we get:
    
    \begin{equation}
        \frac{\partial V(t,x)}{\partial t} + H_{opt} = 0 \nonumber
    \end{equation}
    where:
    
    \begin{equation}
        H_{opt} = \min_{u}(H) = \min_{u}(L+\lambda f) \nonumber
    \end{equation}
    This is called Hamilton-Jacobi-Bellman equation (HJB).
    
    For the stochastic form of this equation, when do Taylor expansion, use Ito's Lemma and take second order.
    
    \section{Avellaneda-Stoikov framework}\label{A2}
    
    The Avellaneda-Stoikov framework provides a foundational approach to modeling the behavior of a market maker operating in a limit order book (LOB). This methodology, introduced in \cite{avellaneda2008high}, combines stochastic control theory with empirical insights from financial markets to derive optimal bid and ask pricing strategies for liquidity providers. Below, we present the key components of the framework, including the underlying assumptions, the derivation of the indifference price, and the solution to the optimization problem.
    
    \subsection*{Empirical Foundations}
    
    The Avellaneda-Stoikov model relies on several empirical observations about financial markets. These include the exponential utility function of the market maker, the diffusion process governing the mid-price, and the Poisson processes describing order arrivals. Specifically, the model assumes:
    \begin{itemize}
        \item The utility function of the market maker is assumed to follow a constant absolute risk aversion (CARA) form:
        \[
            u(x, s, q, t) = -e^{-\gamma (x_t + q_t s_t)}
        \]
        where $x_t$ represents the cash, $s_t$ is the mid-price of the asset, $q_t$ is the inventory level, and $\gamma > 0$ is the risk aversion parameter.
        \item The mid-price $s_t$ evolves as a diffusion process:
        \[
        ds_t = \sigma dW_t,
        \]
        where $W_t$ is standard Brownian motion and $\sigma$ is the volatility of the asset price.
        \item The cash balance $x_t$ changes due to executed trades:
        \[
        dx_t = dN^a_t p^a_t - dN^b_t p^b_t,
        \]
        where $p^a_t = s_t + \delta^a_t$ and $p^b_t = s_t - \delta^b_t$ are the ask and bid prices, respectively, and $N^a_t \sim \text{Poisson}(\lambda^a)$, $N^b_t \sim \text{Poisson}(\lambda^b)$ represent the arrival of sell and buy orders.
    
        \item The inventory process $q_t$ is given by:
        \[
        q_t = N^b_t - N^a_t.
        \]
    \end{itemize}
    
    These assumptions form the foundation for deriving the market maker's optimal strategy, which is represented by the value function:
    
    \begin{equation}
        v(x,s,q,t) = \arg \max_{\delta^a, \delta^b} E_t[-e^{-\gamma (x_T + q_T s_T)}] \label{valuefunction}
    \end{equation}
    
    \subsection*{Indifference Price}
    
    The concept of indifference pricing plays a central role in the Avellaneda-Stoikov framework. The indifference price represents the price at which the market maker is indifferent between holding or not holding an additional unit of inventory. To derive this price, we rewrite the value function $v(x, s, q, t)$ in terms of the reservation price $r$. Starting from the expected utility:
    
    \begin{align}
        E_t[-e^{-\gamma (x+qs_T)}] &= E_t[-e^{-\gamma x} e^{-\gamma q s_T}] 
        \nonumber
        \\
        &= -e^{-\gamma x } E_t[-e^{-\gamma q s_T}] 
        \nonumber
        \\
        &= -e^{-\gamma x} e^{-\gamma q s + \frac{1}{2}(\gamma q)^2 \sigma^2 (T-t)} 
        \nonumber
        \\
        &= -e^{-\gamma x} e^{-\gamma q s} e^{\frac{\gamma^2 q^2 \sigma^2(T+t)}{2}} \label{newvaluefucntionforr}
    \end{align}
    Equaling both functions for $q+1$ and $q-1$ and solve for the bid and ask reservation prices $r^b$ and $r^a$ respectively:
    
    \begin{align}
        v(x-r^b,s,q+1,t) &= v(x,s,q,t) 
        \nonumber
        \\
        -e^{-\gamma(x - r^b)} e^{-\gamma(q + 1)s} e^{\frac{\gamma^2(q + 1)^2 \sigma^2(T - t)}{2}} &= -e^{-\gamma x} e^{-\gamma q s} e^{\frac{\gamma^2 q^2 \sigma^2(T - t)}{2}} 
        \nonumber
        \\
        e^{-\gamma(x - r^b)} e^{-\gamma(q + 1)s} e^{\frac{\gamma^2(q + 1)^2 \sigma^2(T - t)}{2}} &= e^{-\gamma x} e^{-\gamma q s} e^{\frac{\gamma^2 q^2 \sigma^2(T - t)}{2}} 
        \nonumber
        \\
        -\gamma(x - r^b) - \gamma(q + 1)s + \frac{\gamma^2(q + 1)^2 \sigma^2(T - t)}{2} &= -\gamma x - \gamma q s + \frac{\gamma^2 q^2 \sigma^2(T - t)}{2} 
        \nonumber
        \\
        -\gamma x + \gamma r^b - \gamma q s - \gamma s + \frac{\gamma^2(q^2 + 2q + 1) \sigma^2(T - t)}{2} &= -\gamma x - \gamma q s + \frac{\gamma^2 q^2 \sigma^2(T - t)}{2} 
        \nonumber
        \\
        \gamma r^b - \gamma s + \frac{\gamma^2(q^2 + 2q + 1) \sigma^2(T - t)}{2} &= \frac{\gamma^2 q^2 \sigma^2(T - t)}{2} 
        \nonumber
        \\
        \gamma r^b - \gamma s + \frac{\gamma^2(2q + 1) \sigma^2(T - t)}{2} &= 0 
        \nonumber
        \\
        \gamma r^b &= \gamma s - \frac{\gamma^2(2q + 1) \sigma^2(T - t)}{2} 
        \nonumber
        \\
        r^b &= s - \frac{\gamma(2q + 1) \sigma^2(T - t)}{2} 
        \nonumber
        \\
        r^b &= s + (-1 - 2q) \frac{\gamma \sigma^2(T - t)}{2} \label{indiferencepricerb}
    \end{align}
    and similar, for $r^a$ we get:
    \begin{equation}
        r^a = s + (1-2q)\frac{\gamma \sigma^2(T-t)}{2} \label{eq:indiferencepricera}
    \end{equation}
    
    The average of these two prices gives the overall reservation price: 
    \begin{align}
        r &= \frac{r^b + r^a}{2} 
        \nonumber
        \\
         &= \frac{1}{2} \left[ \left( s + (-1 - 2q) \frac{\gamma \sigma^2(T - t)}{2} \right) + \left( s + (1 - 2q) \frac{\gamma \sigma^2(T - t)}{2} \right) \right] 
         \nonumber
         \\
         &= \frac{1}{2} \left[ s + (-1 - 2q) \frac{\gamma \sigma^2(T - t)}{2} + s + (1 - 2q) \frac{\gamma \sigma^2(T - t)}{2} \right]
         \nonumber
         \\
         &= \frac{1}{2} \left[ 2s + \left( (-1 - 2q) + (1 - 2q) \right) \frac{\gamma \sigma^2(T - t)}{2} \right] 
         \nonumber
         \\
         &= \frac{1}{2} \left[ 2s + (-4q) \frac{\gamma \sigma^2(T - t)}{2} \right] 
         \nonumber
         \\
         &= \frac{1}{2} \left[ 2s - 2q \gamma \sigma^2(T - t) \right] 
         \nonumber
         \\
         &= s - q \gamma \sigma^2(T - t) \label{eq:r}
    \end{align}

    \subsection*{Trading Intensity}
    The trading intensity $\lambda(\delta)$ captures the relationship between the spread $\delta$ and the rate of order arrivals. Empirical studies suggest that the price impact $\Delta p$ is proportional to $\ln(Q)$, where $Q$ is the size of market orders. Combining this with the power-law distribution of order sizes $Q(x) \propto x^{-1-\alpha}$ we derive the intensity function:
    \begin{equation*}
        \lambda(\delta) = \Lambda P(\Delta p > \delta)
    \end{equation*}
    where $\Lambda$ is estimated by dividing the total volume traded over a day by the average size of market orders on that day. Substituting the expressions for $\Delta p$:
    \begin{align}
        \lambda(\delta) &= \Lambda P(\ln(Q)>K \delta) 
        \nonumber
        \\
        &= \Lambda P(Q > e^{K \delta}) 
        \nonumber
        \\
        &= \Lambda \int_{e^{K\delta}}^{\infty}x^{-1-\alpha} 
        \nonumber
        \\
        &= \Lambda \frac{e^{-\alpha K\delta}}{\alpha} 
        \nonumber
        \\
        &= A e^{-\kappa \delta} \label{eq:lambda}
    \end{align}
    where $A=\frac{\Lambda}{\alpha}$ and $\kappa=\alpha K$
    
    \subsection*{The Solution}
    
    The market maker's objective is to maximize the expected utility of terminal wealth:
    \begin{equation}
        v(s,x,q,t) = \max_{\delta^a, \delta^b}E_t[-e^{-\gamma(x_T + s_Tq_T)}] \label{eq:valuefunction}
    \end{equation}
    Using dynamic programming principles:
    \begin{equation*}
        v(s,x,q,t) = \max_{\delta^a, \delta^b}E_t[v(s+\Delta s,x+\Delta x, q+\Delta q,t+\Delta t)]
    \end{equation*}
    Assuming $\Delta t$ is small we can expand $v(s+\Delta s,x+\Delta x, q+\Delta q,t+\Delta t)$ using Taylor around (s,x,q,t):
    \begin{align}
        v(s+\Delta s,x+\Delta x, q+\Delta q,t+\Delta t) &\approx v(s,x,q,t) + \frac{\partial v}{\partial t}\Delta t + \frac{\partial v}{\partial s}\Delta s + \frac{\partial v}{\partial x}\Delta x 
        \nonumber
        \\
        &+ \frac{\partial v}{\partial q}\Delta q + \frac{1}{2}\frac{\partial^2 v}{\partial s^2}(\Delta s)^2 + \dots \label{eq:valuefunctiontaylorexp}
    \end{align}
    Considering the expected value $E[v(s+\Delta s, x+\Delta x, q+\Delta q, t+\Delta t)]$, we analyze the following components:
    \begin{itemize}
        \item Price Dynamics: \\
        The price dynamics are given by:
        \[
        \Delta s = \sigma \Delta W_t, \quad \text{where} \quad E_t[\Delta s] = 0 \quad \text{and} \quad E_t[(\Delta s)^2] = \sigma^2 \Delta t.
        \]
        From this, we derive:
        \begin{align}
            E_t\left[\frac{\partial v}{\partial s} \Delta s\right] &= \frac{\partial v}{\partial s} E_t[\Delta s] = 0, 
            \label{eq:expectvaluefirstorder}
            \\
            E_t\left[\frac{1}{2} \frac{\partial^2 v}{\partial s^2} (\Delta s)^2\right] &= \frac{1}{2} \frac{\partial^2 v}{\partial s^2} E_t[(\Delta s)^2] = \frac{1}{2} \sigma^2 \frac{\partial^2 v}{\partial s^2} \Delta t. \label{eq:expectvaluesecondorder}
        \end{align}
    
        \item Cash and Inventory Terms: \\
        For buy orders, the expected change is:
        \begin{equation}
            \lambda^b(\delta^b) \Delta t \left[v(s, x - s + \delta^b, q + 1, t) - v(s, x, q, t)\right] \label{eq:intensitybuy}
        \end{equation}
        For sell orders, the expected change is:
        \begin{equation}
        \lambda^a(\delta^a) \Delta t \left[v(s, x + s + \delta^a, q - 1, t) - v(s, x, q, t)\right] \label{eq:intensitysell}
    \end{equation}
    \end{itemize}
    
    Substituting this \ref{eq:expectvaluefirstorder}, \ref{eq:expectvaluesecondorder}, \ref{eq:intensitybuy} and \ref{eq:intensitysell} in \ref{eq:valuefunctiontaylorexp}, we obtain:
    \begin{align*}
        v(s, x, q, t) = \max_{\delta^a, \delta^b} \bigg[ &v(s, x, q, t) + \frac{\partial v}{\partial t} \Delta t + \frac{1}{2} \sigma^2 \frac{\partial^2 v}{\partial s^2} \Delta t \\
        &+ \lambda^b(\delta^b) \Delta t \big(v(s, x - s + \delta^b, q + 1, t) - v(s, x, q, t)\big) \\
        &+ \lambda^a(\delta^a) \Delta t \big(v(s, x + s + \delta^a, q - 1, t) - v(s, x, q, t)\big) \bigg]
    \end{align*}
    
    By focusing on the maximization, we can cancel \( v(s, x, q, t) \) and divide through by \( \Delta t \), yielding:
    \begin{align}
        \frac{\partial v}{\partial t} + \frac{1}{2} \sigma^2 \frac{\partial^2 v}{\partial s^2} &+ \max_{\delta^b} \big[\lambda^b(\delta^b) \big(v(s, x - s + \delta^b, q + 1, t) - v(s, x, q, t)\big)\big] 
        \nonumber
        \\
        &+ \max_{\delta^a} \big[\lambda^a(\delta^a) \big(v(s, x + s + \delta^a, q - 1, t) - v(s, x, q, t)\big)\big] = 0 \label{eq:hjb}
    \end{align}
    with the terminal condition:
    \[
    v(s, x, q, T) = -\exp{(-\gamma (x + q s))}
    \]
    
    Due to the choice of exponential utility, we can simplify the problem using the ansatz:
    \begin{equation*}
        v(s, x, q, t) = -\exp{(-\gamma x)} \exp{(-\gamma \theta(s, q, t))}
    \end{equation*}
    where:
    \begin{itemize}
        \item \( \gamma \) is the risk aversion parameter, and
        \item \( \theta(s, q, t) \) is a function to be determined.
    \end{itemize}
    This ansatz separates the dependence on \( x \) (cash) from the dependence on \( s \) (price), \( q \) (inventory), and \( t \) (time). The next step is to compute the partial derivatives of \( v(s, x, q, t) \) with respect to \( t \), \( s \), and \( x \), as they appear in the HJB equation:
    
    \begin{align}
        \frac{\partial v}{\partial t} &= \gamma \frac{\partial \theta}{\partial t} v 
        \label{eq:partialtheta}
        \\
        \frac{1}{2} \sigma^2 \frac{\partial^2 v}{\partial s^2} &= \frac{1}{2} \sigma^2 \left( \gamma \frac{\partial^2 \theta}{\partial s^2} v - \gamma^2 \left( \frac{\partial \theta}{\partial s} \right)^2 v \right)
        \label{eq:partialsecondordertheta}
        \\
        v(s, x - s + \delta^b, q + 1, t) &= -\exp{(-\gamma (x - s + \delta^b))} \exp{(-\gamma \theta(s, q + 1, t))} 
        \label{eq:valuecuntiontheta}
        \\
        v(s, x - s + \delta^b, q + 1, t) - v(s, x, q, t) &= v(s, x, q, t) \left[ e^{\gamma (s - \delta^b)} e^{-\gamma (\theta(s, q + 1, t) - \theta(s, q, t))} - 1 \right] \label{eq:changetheta}
    \end{align}
    Substituting \ref{eq:partialtheta}, \ref{eq:partialsecondordertheta}, \ref{eq:valuecuntiontheta} and \ref{eq:changetheta} into the HJB equation \ref{eq:hjb} and dividing through by \( v(s, x, q, t) \), we obtain:
    \begin{align*}
        \gamma \frac{\partial \theta}{\partial t} + \frac{1}{2} \sigma^2 \left( \gamma \frac{\partial^2 \theta}{\partial s^2} - \gamma^2 \left( \frac{\partial \theta}{\partial s} \right)^2 \right) &+ \max_{\delta^b} \lambda^b(\delta^b) \left[ e^{\gamma (s - \delta^b)} e^{-\gamma (\theta(s, q + 1, t) - \theta(s, q, t))} - 1 \right]\\
        &+ \max_{\delta^a} \lambda^a(\delta^a) \left[ e^{-\gamma (s + \delta^a)} e^{-\gamma (\theta(s, q - 1, t) - \theta(s, q, t))} - 1 \right] = 0
    \end{align*}
    dividing by \( \gamma \) we simplify to:
    \begin{align*}
        \frac{\partial \theta}{\partial t} + \frac{1}{2} \sigma^2 \left( \frac{\partial^2 \theta}{\partial s^2} - \gamma \left( \frac{\partial \theta}{\partial s} \right)^2 \right) &+ \max_{\delta^b} \frac{\lambda^b(\delta^b)}{\gamma} \left[ e^{\gamma (s - \delta^b)} e^{-\gamma (\theta(s, q + 1, t) - \theta(s, q, t))} - 1 \right]\\
        &+ \max_{\delta^a} \frac{\lambda^a(\delta^a)}{\gamma} \left[ e^{-\gamma (s + \delta^a)} e^{-\gamma (\theta(s, q - 1, t) - \theta(s, q, t))} - 1 \right] = 0
    \end{align*}
    using the definition of reservation prices:
    \begin{itemize}
        \item \( r^b = \theta(s, q + 1, t) - \theta(s, q, t) \),
        \item \( r^a = \theta(s, q - 1, t) - \theta(s, q, t) \),
    \end{itemize}
    we rewrite the equation as:
    \begin{align}
        \frac{\partial \theta}{\partial t} + \frac{1}{2} \sigma^2 \frac{\partial^2 \theta}{\partial s^2} - \frac{1}{2} \sigma^2 \gamma \left( \frac{\partial \theta}{\partial s} \right)^2 &+ \max_{\delta^b} \frac{\lambda^b(\delta^b)}{\gamma} \left[ 1 - e^{\gamma (s - \delta^b - r^b)} \right]
        \nonumber
        \\
        &+ \max_{\delta^a} \frac{\lambda^a(\delta^a)}{\gamma} \left[ 1 - e^{-\gamma (s + \delta^a - r^a)} \right] = 0 \label{eq:newhjbtheta}
    \end{align}
    
    To find the optimal \(\delta^b\) and \(\delta^a\) we maximize the terms:
    \[
    \max_{\delta^b} \frac{\lambda^b(\delta^b)}{\gamma} \left[ 1 - e^{\gamma (s - \delta^b - r^b)} \right]
    \]
    \[
    \max_{\delta^a} \frac{\lambda^a(\delta^a)}{\gamma} \left[ 1 - e^{-\gamma (s + \delta^a - r^a)} \right]
    \]
    This involves taking the derivative of each term with respect to \(\delta^b\) and \(\delta^a\), setting the derivative to zero, and solving for \(\delta^b\) and \(\delta^a\):
    
    \begin{align}
        \frac{\partial f}{\partial \delta^b} = \frac{1}{\gamma} \left[ \frac{\partial \lambda^b}{\partial \delta^b} \left( 1 - e^{\gamma (s - \delta^b - r^b)} \right) + \lambda^b(\delta^b) \cdot \gamma e^{\gamma (s - \delta^b - r^b)} \right] &= 0 
        \nonumber
        \\
        \frac{\partial \lambda^b}{\partial \delta^b} \left( 1 - e^{\gamma (s - \delta^b - r^b)} \right) + \gamma \lambda^b(\delta^b) e^{\gamma (s - \delta^b - r^b)} &= 0 
        \nonumber
        \\
        \frac{\partial \lambda^b}{\partial \delta^b} \left( 1 - e^{\gamma (s - \delta^b - r^b)} \right) &= -\gamma \lambda^b(\delta^b) e^{\gamma (s - \delta^b - r^b)} 
        \nonumber
        \\
        \frac{\partial \lambda^b}{\partial \delta^b} \left( \frac{1}{e^{\gamma (s - \delta^b - r^b)}} - 1 \right) &= -\gamma \lambda^b(\delta^b) 
        \nonumber
        \\
        \frac{\partial \lambda^b}{\partial \delta^b} \left( e^{-\gamma (s - \delta^b - r^b)} - 1 \right) &= -\gamma \lambda^b(\delta^b) 
        \nonumber
        \\
        e^{-\gamma (s - \delta^b - r^b)} &= 1 - \gamma \frac{\lambda^b(\delta^b)}{\frac{\partial \lambda^b}{\partial \delta^b}} 
        \nonumber
        \\
        -\gamma (s - \delta^b - r^b) &= \ln \left( 1 - \gamma \frac{\lambda^b(\delta^b)}{\frac{\partial \lambda^b}{\partial \delta^b}} \right) 
        \nonumber
        \\
        s - r^b &= \delta^b - \frac{1}{\gamma} \ln \left( 1 - \gamma \frac{\lambda^b(\delta^b)}{\frac{\partial \lambda^b}{\partial \delta^b}} \right) \label{eq:optimalrb}
    \end{align}
    Similar procedure for the term involving \(\delta^a\):
    \begin{equation}
        r^a - s = \delta^a - \frac{1}{\gamma} \ln \left( 1 - \gamma \frac{\lambda^a(\delta^a)}{\frac{\partial \lambda^a}{\partial \delta^a}} \right) \label{eq:optimalra}
    \end{equation}
    
    Now, using the reservation price \ref{eq:r} and the exponential arrival rate \ref{eq:lambda} in the optimal distances \ref{eq:optimalrb} and \ref{eq:optimalra} we obtain:
    
    \begin{align}
        s - [s-(q+1)\gamma \sigma^2(T-t)] &= \delta^b - \frac{1}{\gamma}\ln(1-\gamma\frac{\lambda(\delta^b)}{\frac{\partial \lambda}{\partial \delta}}) 
        \nonumber
        \\
        (q+1)\gamma \sigma^2(T-t) &= \delta^b - \frac{1}{\gamma}\ln(1-\gamma \frac{\lambda}{-\kappa \lambda}) 
        \nonumber
        \\
        (q+1)\gamma \sigma^2(T-t) &= \delta^b - \frac{1}{\gamma}\ln(1+\frac{\gamma}{\kappa}) 
        \nonumber
        \\
        \delta^b &= q \gamma \sigma^2(T-t) + \frac{1}{\gamma}\ln(1+\frac{\gamma}{\kappa}) \label{eq:deltab}
    \end{align}
    and similar for $\delta^a$ we get:
    \begin{equation}
        \delta^a = -q \gamma \sigma^2(T-t) + \frac{1}{\gamma}\ln(1+\frac{\gamma}{\kappa}) \label{eq:deltaa}
    \end{equation}
    
    $(q+1)$ and $(q-1)$ term is replaced with $q$ because the optimal distance $\delta$ is expressed in terms of the current inventory level and not the future inventory level. This strategy therefore quotes a fixed spread of:
    \begin{equation*}
        \delta^b + \delta^a = \frac{2}{\gamma}\ln(1+\frac{\gamma}{\kappa})
    \end{equation*}
    
    \section{Informal Settings}\label{A3}
    
    In this appendix, we derive the optimal bid and ask distances ($\delta^b$ and $\delta^a$) for a market maker in an informal market setting. We use the same framework of Avellaneda-Stoikov and begin with the equations derived in Appendix \ref{A2}:
    \begin{align*}
        s - r^b &= \delta^b - \frac{1}{\gamma} \ln \left( 1 - \gamma \frac{\lambda^b(\delta^b)}{\frac{\partial \lambda^b}{\partial \delta^b}} \right) \\
        r^a - s &= \delta^a - \frac{1}{\gamma} \ln \left( 1 - \gamma \frac{\lambda^a(\delta^a)}{\frac{\partial \lambda^a}{\partial \delta^a}} \right) \tag{C1} \label{eq:coptimaldistance}
    \end{align*}
    Under the assumption that the intensity of market orders follows an exponential decay $\lambda(\delta) = Ae^{-\kappa \delta}$, Avellaneda-Stoikov derived the following optimal distances:
    \begin{align*}
        \delta^b &= q \gamma \sigma^2(T-t) + \frac{1}{\gamma}\ln(1+\frac{\gamma}{\kappa}) \\
        \delta^a &= -q \gamma \sigma^2(T-t) + \frac{1}{\gamma}\ln(1+\frac{\gamma}{\kappa})
    \end{align*}
    
    However, our empirical study in this informal market reveals that while the price impact function $\Delta p$ exhibits a logarithmic dependence on the market order size ($\Delta p \propto ln(Q)$, Figure ),  the density of market order sizes does not follow a power law. Instead, it follows an exponential distribution: $Q(x) \propto e^{-\alpha x}$. Consequently, by recalculating the intensity $\lambda$ using this updated function, we obtain:
    
    \begin{align*}
        \lambda(\delta)&=\Lambda P(\Delta P > \delta) \\
        &=\Lambda P(ln(Q) > K \delta) \\
        &=\Lambda P(Q > e^{K \delta}) \\
        &=\Lambda \int_{e^{K \delta}}^{\infty}e^{-\alpha x}dx \\
        &=\frac{\Lambda}{\alpha}e^{-\alpha e^{K \delta}} \\
        &=Ae^{-\alpha e^{K \delta}}
    \end{align*}
    Then, we need to calculete a new optimal distances taken acount that:
    \begin{align*}
        \frac{\partial \lambda}{\partial \delta} &= Ae^{-ae^{K\delta}}(-ae^{K\delta})K \\
        &= -AaKe^{K\delta}e^{-aK\delta}
    \end{align*}
    then, the rate $\frac{\lambda(\delta)}{\frac{\partial \lambda}{\partial \delta}}$ is:
    \begin{align*}
        \frac{\lambda(\delta)}{\frac{\partial \lambda}{\partial \delta}} &= \frac{Ae^{-ae^{K\delta}}}{-AaKe^{K\delta}e^{-aK\delta}} \\
        &= -\frac{1}{aKe^{K\delta}} \tag{C2} \label{eq:newlambda}
    \end{align*}
    Substituing \ref{eq:newlambda} in the optimal distance equations \ref{eq:coptimaldistance}:
    \begin{align*}
        s - [s-(q+1)\gamma \sigma^2(T-t)] &= \delta^b - \frac{1}{\gamma}\ln(1-\gamma(-\frac{1}{aKe^{K\delta}})) \\
        (q+1)\gamma \sigma^2(T-t) &= \delta^b - \frac{1}{\gamma}\ln(1+\frac{\gamma}{aKe^{K\delta}}) \\
        \delta^b &= (q+1)\gamma \sigma^2(T-t) + \frac{1}{\gamma}\ln(1+\frac{\gamma}{aKe^{K\delta}}) \\
        \delta^b &= q\gamma \sigma^2(T-t) + \frac{1}{\gamma}\ln(1+\frac{\gamma}{aKe^{K\delta}})
    \end{align*}
    similar for $\delta^a$ we get:
    \begin{equation*}
        \delta^a = -q \gamma \sigma^2(T-t) + \frac{1}{\gamma}\ln(1+\frac{\gamma}{aKe^{K\delta}})
    \end{equation*}
    
    \section{New price algorithm}\label{A4}
    
    The algorithm iterates through each time period $t$ and calculates the bid and ask prices using the Avellaneda-Stoikov model, denoted as $S^b_t$ and $S^a_t$ respectively. These prices serve as references for the market maker's actions, which are determined by the type of order and its price relative to the current bid and ask prices.
    
    If an order is a buy order ($\epsilon_x = +1$) and $p_x > s^a_t$, the market maker sell at $S^a_t$. Conversely, for a sell order $\epsilon=-1$, the market maker buy at $s^b_t$ if $p_x < s^b_t$. After processing all orders for a given time period, the algorithm updates the volumes on the bid and ask sides of the LOB and calculates a new price using the formula
    \begin{equation}
        \hat{s}_{t+1} = \hat{s}_t + \xi_t(Q^b_t - Q^a_t) + \epsilon_t
    \end{equation}
    finally, the algorithm adjusts each order price $p_x$ for the next time period $t+1$ using $\hat{s}$ and $s$.

    \begin{algorithmic}[1]
        \State $s_0 \gets s$
        \For{$t \in T$}
            \State Calculate $s^a_t$, $s^b_t$ $\gets$ Avellaneda-Stoikov
            \For{$x \in t$}
                \If{$\epsilon_x = -1$ and $x_p < s^b_t$}
                    \State MM buy $x$ in $s^b_t$
                \ElsIf{$\epsilon_x = 1$ and $s^a_t < x_p$} 
                    \State MM sell in $s^a_t$
                \EndIf
            \EndFor
            \State Update $Q^b_{t+1} = \sum_{i=1}^{b}v^b_i$
            \State Update $Q^a_{t+1} = \sum_{i=1}^{a}v^a_i$
            \State Calculate $\hat{s}_{t+1} = s_t + \xi_t \cdot (Q^b_t - Q^a_t)$ 
            \State Update $x_p \in t+1$:
            \[
            \hat{p_x}_{t+1} = \frac{p_x \cdot \hat{s}_{USD}}{s_{USD}}
            \]
        \EndFor
    \end{algorithmic}
    
    This logic, although simple, allows us to estimate a new USD price $\hat{s}$ under the influence of a monopolistic market maker.

    \end{appendices}

\bibliographystyle{plain}
\bibliography{references}  

\end{document}